\documentclass[aps,preprint,superscriptaddress,prb,amsmath]
{revtex4}
\usepackage{txfonts}
\usepackage{color}
\usepackage[latin9]{inputenc}
\usepackage{amssymb}
\usepackage{graphicx}
\usepackage{dcolumn}
\usepackage{bm}
\usepackage{hyperref}
\usepackage{color}

\begin{document}
\title{Hybrid single-pair charge-2 Weyl semimetals}
\author{P. Zhou}
\email{zhoupan71234@xtu.edu.cn}
\affiliation{Hunan Provincial Key laboratory of Thin Film Materials and Devices, School of Material Sciences and Engineering, Xiangtan University, Xiangtan 411105, China}
\author{Y. Z. Hu}
\affiliation{Hunan Provincial Key laboratory of Thin Film Materials and Devices, School of Material Sciences and Engineering, Xiangtan University, Xiangtan 411105, China}
\author{B. R. Pan}
\affiliation{Key Laboratory of Low-dimensional Materials and Application Technology, School of Material Sciences and Engineering, Xiangtan University, Xiangtan 411105, China}
\author{F. F. Huang}
\affiliation{Key Laboratory of Low-dimensional Materials and Application Technology, School of Material Sciences and Engineering, Xiangtan University, Xiangtan 411105, China}
\author{W. Q. Li}
\affiliation{Key Laboratory of Low-dimensional Materials and Application Technology, School of Material Sciences and Engineering, Xiangtan University, Xiangtan 411105, China}
\author{Z. S. Ma}
\affiliation{Hunan Provincial Key laboratory of Thin Film Materials and Devices, School of Material Sciences and Engineering, Xiangtan University, Xiangtan 411105, China}
\author{L. Z. Sun}
\email{lzsun@xtu.edu.cn}
\affiliation{Hunan Provincial Key laboratory of Thin Film Materials and Devices, School of Material Sciences and Engineering, Xiangtan University, Xiangtan 411105, China}
\date{\today}
%
%
\begin{abstract}
Intuitively, the dispersion characteristics of Weyl nodes with opposite charges in single-pair charge-2 Weyl semimetals are the same, quadratic or linear. We theoretically predicted that single-pair hybrid charge-2 Weyl semimetals (the nodes with opposite charges show quadratic Weyl and linear charge-2 Dirac characteristics, respectively) can be protected by specific nonsymmorphic symmetries in spinless systems. Moreover, the symmetries force the pair of Weyl points locate at the center and corners of the first Brillouin zone (FBZ), respectively. Consequently, nontrivial surface states run through the entire FBZ of the system fascinating for future experimental detection and device applications. The hybrid phase is further verified with the help of first-principles calculations for the phonon states in realistic material of Na$_2$Zn$_2$O$_3$. The new phase will not only broaden the understanding of the Weyl semimetals, but also provide an interesting platform to investigate the interaction between the two types of Weyl fermions with different dispersions.\\
\end{abstract}
\maketitle
\indent In high-energy physics, Weyl fermions\cite{w1,w2,w3,w4,w5,w6} are relativistic massless particles that are restricted by the Poincare symmetry. The low-energy quasiparticle excitations of Weyl fermions in condensed matters protected by crystalline symmetries are alternative platforms to study the novel physics of the elementary particles\cite{gama,W_S_2,W_S_3,W_S_4}. The Weyl fermions are monopoles of Berry curvature in the Brillouin zone (BZ) and the charges associated with the monopoles are known as their chirality\cite{Weyl_chiral_1,Weyl_chiral_2,chiral_1,chiral_2}. In crystalline materials, the Bloch function enriches the chiral charges to be $\pm$1\cite{Chern_number1_1,Chern_number1_2,Chern_number1_3,Chern_number1_4,Chern_number1_5,Chern_number1_6,Chern_number1_7}, $\pm$2\cite{prl2011,Chern_number2_2,Chern_number2_3,Chern_number2_4,Chern_number2_5,Chern_number2_6,Chern_number2_7}, $\pm$3\cite{Chern_number_3_1,Chern_number_3_2}, and even $\pm$4\cite{Chern_number_4_1,Chern_number_4_2,Chern_number_4_3,Chern_number_4_4}, relying on the degree of degeneracy of the Weyl points and band dispersion around the Weyl nodes. In comparison with the conventional nodes\cite{c1,c2,c3,c4,c5} with the charges of $\pm$1, the nodes with higher charge\cite{uc1,uc2,uc3,uc4,uc5} are fascinating since the chiral charges determine the number of chiral edge states. Consequently, the Weyl points with higher monopole charges will result in intensified Fermi arcs. The anomalous Hall effect\cite{Anomalous1,Anomalous2,Chiral_Anomaly_3}, as well as the effects linked to the chiral anomaly such as negative magnetic resistance\cite{Chiral_Anomaly_1,uc4}, will be strengthened with the increase in the monopole charge. Furthermore, the intrinsic anisotropy of quadratic or cubic dispersion of  higher charge Weyl cones will result in novel correlation phenomena. Therefore, Weyl semimetals with higher charge receive great attention, recently.\\
\begin{figure}
	\center
	\includegraphics[trim={0.0in 0.0in 0.0in 0.0in},clip,width=4.5in]{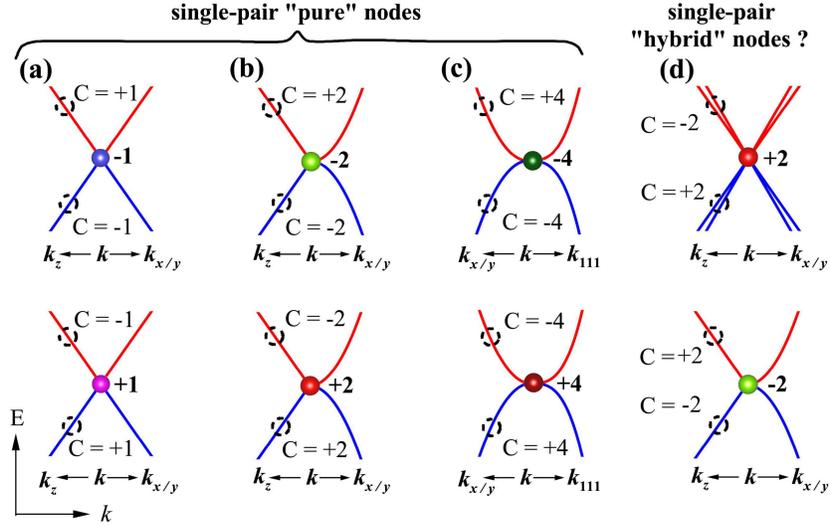}
	\caption{Four types of single-pair Weyl nodes. (a) Single-pair "pure" nodes with a chiral charge of $\pm$1 and linear dispersion in all directions\cite{MnBi2Te4,Chern_number1_7}. (b) Single-pair "pure" nodes with a chiral charge of $\pm$2\cite{NM_single-pair,a_system_2}. There are a quadratic dispersion in specific plane and linear dispersion in the perpendicular orientations to the plane. (c) Single-pair "pure" nodes with a chiral charge of $\pm$4\cite{Chern_number_4_2}. The nodes show a $\textbf{\emph{k}}^3$-type dispersion splitting along one direction and quadratic dispersion splitting along other directions. (d) Single-pair "hybrid" nodes, including a double-Weyl point (twofold degenerate point) with quadratic dispersion and a charge-2 Dirac point (fourfold-degenerate point) with linear dispersion, respectively.}\label{fig1}
\end{figure}
\indent As for the studies on Weyl semimetals, one of the critical obstacles is that there are usually too many pairs of Weyl points around the Fermi level in condensed matters, for example 12 pairs of Weyl nodes in TaAs\cite{W_S_2,w3,W_S_3,TaAs_4}, which will complicate their experimental detections and device applications. Thus, to find Weyl semimetals with fewer pairs of Weyl nodes arouses much interest. According to the Nielsen-Ninomiya no-go theorem\cite{no-go_theorem_1,no-go_theorem_2}, the total chirality in the entire three-dimensional BZ of a Weyl semimetal must be zero. Therefore, theoretically, the minimum number of Weyl nodes in a condensed matter is single-pair. This kind of system will be helpful for us to capture the main physical phenomenon and the associated topological features through simplifying the theoretical analysis and related experimental results. Very recently, several Weyl semimetals with single-pair Weyl nodes have been reported\cite{prl2011,MnBi2Te4,EuCd2As2,Chern_number1_7,NM_single-pair, Chern_number_4_2,a_system_2}. The chiral charges of them can be $\pm$1, $\pm$2, and $\pm$4. Up to now, the two Weyl nodes with opposite charges in single-pair Weyl semimetals are all the same dispersion characteristics, as shown in Fig. 1(a)-1(c), we call them ``pure'' type single-pair Weyl semimetals. As for charge-2 single-pair Weyl semimetals, considering half-filling and excluding spin-1 Weyl node,  there are two kinds of typical band dispersion for charge-2 Weyl nodes: quadratic Weyl and charge-2 Dirac nodes\cite{Double-Weyl_1,Double-Weyl_2,Double-Weyl_3,Double-Weyl_4,Chern_number2_5}. The quadratic Weyl node (also called double-Weyl node) is twofold degenerate with quadratic dispersion along specific high symmetry lines, as shown in Fig. \ref{fig1}(b). Whereas the charge-2 Dirac node is composed of degenerate Weyl nodes with same chiral charges. Until now,  single-pair charge $\pm$2 Weyl semimetals previously reported are all pure type, as shown in Fig. \ref{fig1}(b). An interesting issue is that can the symmetries of condensed matter enforce the single-pair "hybrid" type charge-2 Weyl semimetal? Namely, the nodes with opposite charges show quadratic Weyl and linear charge-2 Dirac characteristics, respectively, as shown in Fig. \ref{fig1}(d). Such new phase will not only broaden the understanding of the Weyl semimetals, but also provide an interesting platform to investigate the interaction between the two kinds of Weyl fermions with different dispersions.\\
\indent In present letter, using the theory of elementary band representation\cite{EBR1,EBR2} (EBR) and considering the symmetries of the Bloch states at high symmetry point and their combinatorics between different wavevectors\cite{slager1, slager2}, we theoretically predicted that hybrid single-pair charge-2 Weyl semimetal can be protected by specific nonsymmorphic space groups, such as $P4_{1}2_{1}2$ and $P4_32_12$, in spinless systems. Moreover, the symmetries enforce the pair of Weyl points locate at the center and corners of the FBZ, respectively, that any two-dimensional plane in the reciprocal space (except for those containing the Weyl points) is topologically nontrivial with nonzero Chern number (1 in $k_z$ $>$ 0 and -1 in $k_z$ $<$ 0 in FBZ). Consequently, nontrivial surface states run through the entire surface FBZ of the system due to the unconventional Chern number distribution. Inspired by the theoretical results, we further verified the new charge-$\pm$2 Weyl phase in the optical phonon states of realistic material Na$_2$Zn$_2$O$_3$ with the help of first-principles calculations, which is fascinating for future experimental detection and device applications.\\
\begin{figure}
	\center
	\includegraphics[trim={0.0in 2.5in 0.0in 2.0in},clip,width=5.35in]{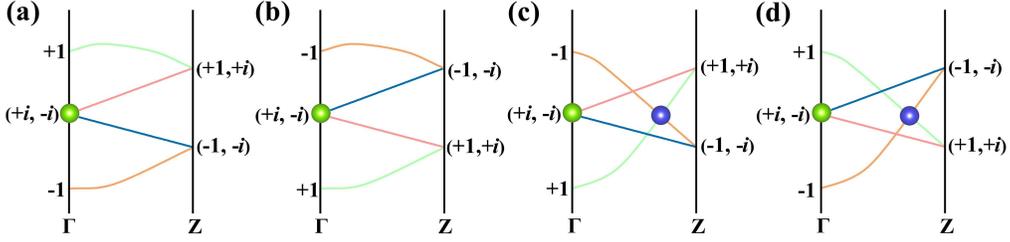}
	\caption{Four potential band structures forming double-Weyl node along high-symmetry line $\Gamma$-Z, where +1, -1, $+i$, and $-i$ are the eigenvalues of the operation ${C}_{4}$. The time-reversal symmetry results in the degeneracy of the eigenstates at the $\Gamma$ point and the degeneracy point located at the Fermi level if half-filling is taken into account. The purple dots show the conventional Weyl points with linear dispersion in all directions, while the green points show the double-Weyl nodes with quadratic dispersion in the k$_z$ = 0 plane and linear dispersion along $\Gamma$-Z.}\label{fig2}
\end{figure}
\begin{figure}
	\centering
	\includegraphics[trim={0.0in 0.0in 0.0in 0.0in},clip,width=4.4in]{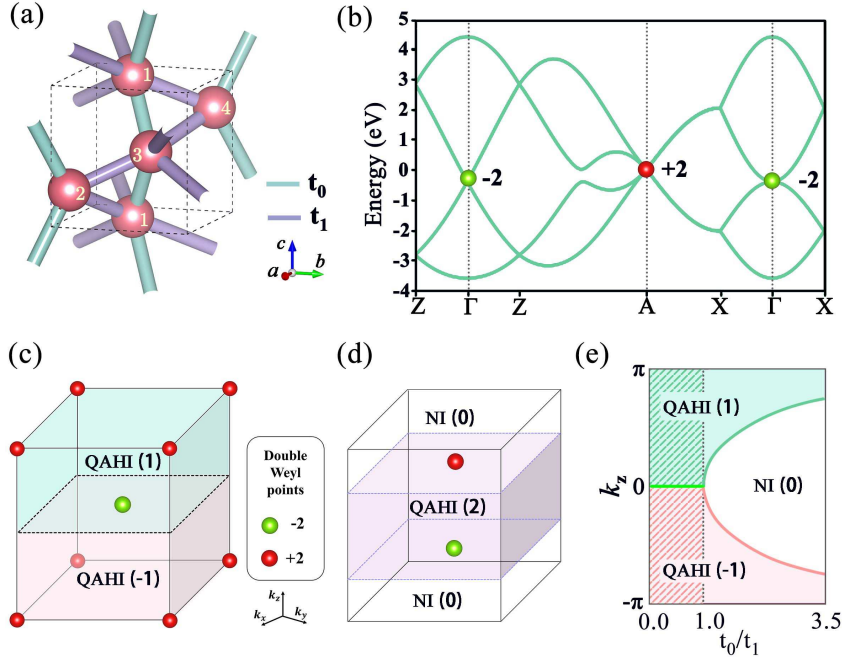}
	\caption{(a) The crystal structure adopted in the TB model. The bonds with sapphire and purple denote the nearest-neighbor hopping and next nearest-neighbor hopping terms, respectively. The fractional coordinates of the four sites in the unit cell are (0.40, 0.40, 0.00), (0.90, 0.10, 0.25), (0.60, 0.60, 0.50), and (0.10, 0.90, 0.75), respectively. (b) Energy band structure from TB model and the $\pm$ 2 are the Chern numbers of the Weyl Points. (c) and (d) are the distribution of charge-2 Weyl points in our model and conventional charge-2 Weyl semimetal\cite{prl2011}, respectively. The red and green balls indicate the Weyl points with Chern number +2, and -2, respectively. (e) Phase diagram of the TB model with different $t_0$/$t_1$. The QAHI phases with Chern number +1 [QAHI(1)] and -1 [QAHI(-1)] are shaded in green, and red, respectively. The white area indicates the normal insulator (NI).}	\label{fig3}
\end{figure}
\indent In present work, we focus on the three-dimensional (3D) Weyl semimetal in class AI (spinless and time-reversal symmetric) with additional crystalline symmetries. The single-valued representations can be used to characterize the symmetry features of this class of systems. Concretely, they can be used to describe spinless electrons, phonons of crystalline materials or some artificial systems. By examining the little cogroup irreducible representations (irreps) of all 230 space groups, we find the single-pair of hybrid charge-2 Weyl nodes can be realized in the space groups of $P4_{1}2_{1}2$ and $P4_32_12$, and the details will be discussed below. Considering the similar characteristics, we concentrate on the space group $P4_32_12$ (No. 96). Under time-reversal symmetry, there are two EBRs in this space group and they can be induced from the Wyckoff position 4a with the local site point group $C_2$. At the $A$ point, only two conjugated two-order little cogroup irreps exist: $A_1\oplus A_2$, indicating that the bands at this point must be fourfold-degenerate. The fourfold degeneracy can be derived from the anticommutation between S$_{2x}$ and S$_{2y}$, as well as their commutation relation with time-reversal operator $T$. A \textit{\textbf{k}}$ \cdot $\textit{\textbf{p}} model analysis reveals that the dispersion is linear along all directions (see details in Supplemental materials\cite{sup}). Therefore, a double Weyl node will form at the high symmetry point. $\Gamma$-Z is a $C_4$ rotation invariant path and the states on this path can be written as the eigenstates of $C_4$. As mentioned above, the set of EBRs can be induced from Wyckoff position 4a. Considering the eigenvalues at $\Gamma$ and Z as listed in Tab. S1 as well as the compatibility relations, there are 12 types of energy band connections as shown in Fig. 2 and Fig. S1. For all the cases, Weyl points are formed under half-filling condition. However, only four types of energy band connections can produce charge-2 Weyl node, as shown in Fig. 2. Among them, only two cases as shown in Figs. 2(a) and 2(b) can fulfill the single-pair charge-2 Weyl semimetals. Considering the similarity of the two charge-2 Weyl nodes, we take the case as shown in Fig. 2(a) as an example. At high symmetry point $\Gamma$ with the wave vector $q_{\Gamma} = (0,0,0)$, the Hamiltonian is constrained by the point group $D_{4h}$. Considering the time-reversal invariant of $\Gamma$ point, the double-degenerate states must have the $C_4$ eigenvalues $\pm{i}$, then the $u_c$/$u_v$ = -1, which indicates the bands is quadratic in $q_{x,y}$ with the monopole charge $\pm$2\cite{gama}. If the charge-2 Dirac point at  $A$ point and quadratic Weyl point around the $\Gamma$ point share the same "occupation", the system will feature hybrid single-pair Weyl nodes.\\
\begin{figure}
	\includegraphics[trim={0.0in 0.0in 0.0in 0.0in},clip,width=5.4in]{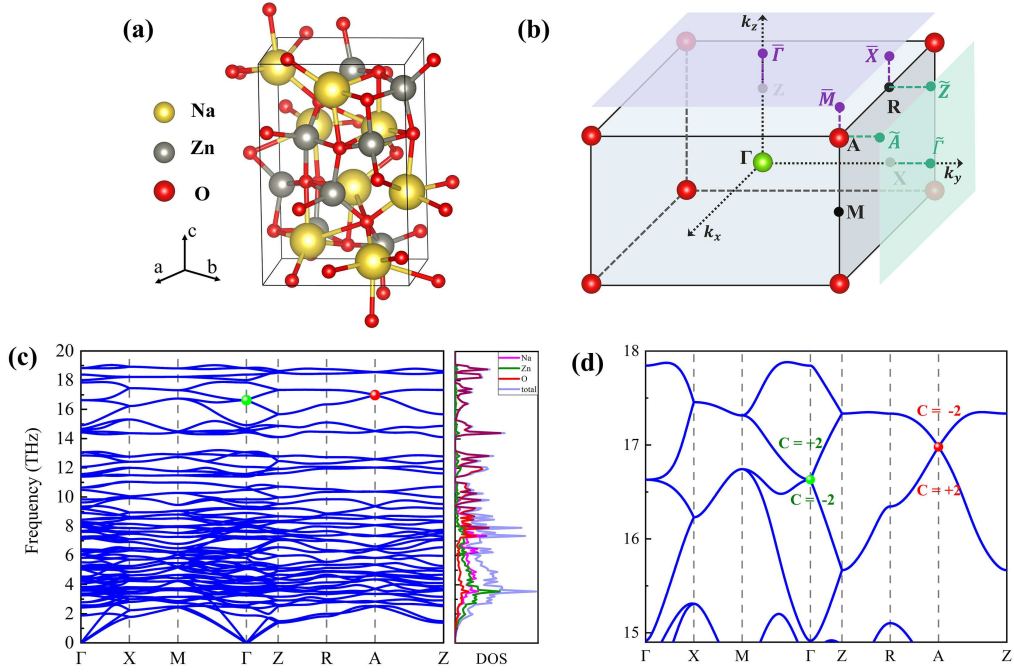}\\
	\caption{(a) The crystal structure of Na$_ 2$Zn$_ 2$O$_ 3$, where the yellow, gray and red spheres represent the Na, Zn and O atoms, respectively. (b) The bulk BZ and the projected surface BZ for the (001) and (010) surfaces, respectively. The charge-2 points with Chern number -2, and +2 are schematically shown by the green and red spheres, respectively. (c) Calculated phonon dispersion and DOS. The zoom in phonon spectra in the range of 14.9 to 18.0 THz.  (d) The Chern numbers of such nontrivial phonon branches are illustrated around the $\Gamma$ and A points. }\label{fig4}
\end{figure}
\indent To further confirm and investigate the general properties of a system with a single-pair of hybrid charge-2 nodes, we construct a four-site TB model as shown in Fig. 3(a). In the model, a tetragonal lattice with four sites per unit cell is adopted. The four sites locate at the Wyckoff 4a of the space group $P4_{3}2_{1}2$. The fourfold screw symmetry S$_{4z}$ := ($x$, $y$, $z$) $\rightarrow$ ($y + \frac{1}{2}$, $-x + \frac{1}{2}$, $z + \frac{1}{4}$) makes the atoms going from 1 $\rightarrow$ 2 $\rightarrow$ 3 $\rightarrow$ 4, as shown in Fig. 3(a). The Hamiltonian with only nearest-neighbor and next-nearest-neighbor hopping terms can be written as
\begin{equation}\label{eq1}
	\begin{aligned}
		H = &\sum_{\left \langle i, j\right \rangle}t_{0}(a_{i}^{\dagger}c_{j}+b_{i}^{\dagger}d_{j})\\
		&+ \sum_{\left \langle \left \langle i, j\right \rangle \right \rangle}t_{1}(a_{i}^{\dagger}b_{j}+a_{i}^{\dagger}d_{j}+ b_{i}^{\dagger}c_{j}+ c_{i}^{\dagger}d_{j})  + H.c.,
	\end{aligned}
\end{equation}
where $a^{\dagger}(a)$, $b^{\dagger}(b)$, $c^{\dagger}(c)$, and $d^{\dagger}(d)$ are the creation (annihilation) operators on sites 1 to 4, respectively. $t_{0}$ and $t_{1}$ are the nearest-neighbor and next nearest-neighbor hopping parameters, respectively. The band structures derived from the TB model with $ t_{0}$ = 0.4 and $t_{1} = 1.0$ (the evolution of the band structures according to the parameters will be discussed below) are shown in Fig. 2(b). The results indicate that, at half-filling, the energy bands host a doubly-degenerate point at the $\Gamma$ point with linear dispersion along $\Gamma$-Z high symmetry line and quadratic along $\Gamma$-X high symmetry line, whereas a linear fourfold point forms at the A point. The chiral charge or Chern numbers of them are -2, and +2, respectively. The symmetry enforced distribution of the two types of charge-2 Weyl points is shown in Fig. 3(c). Generally, for a Weyl semimetal, quantum anomalous Hall insulator (QAHI) phase is restricted in the space of BZ between the two Weyl points\cite{prl2011}. For example, as for charge-2 semimetals, the Chern number of QAHI phase between the Weyl points is 2, denoted as QAHI(2) as shown in Fig. 3(d). The results of the hybrid single-pair Weyl semimetal show that the QAHI phase distribute throughout the whole 3D BZ with Chern numbers of +1 [QAHI(1)] and -1 [QAHI(-1)] for $k_z> 0$ and k$_{z}$ $<$ 0, respectively, as shown in Fig. 3(c). The result is similar to the case of higher-order Weyl semimetals\cite{high1}, the different Chern numbers on either sides of $k_z$ = 0 plane are caused by the quadratic Weyl point at the $\Gamma$. Consequently, nontrivial surface states run through the entire surface FBZ of the system due to the unconventional Chern number distribution, which will be favorite for experimental detection and device applications.\\
\indent To investigate the appearance condition of the hybrid single-pair Weyl semimetal phase, the phase diagram in function of $ t_{0}/ t_{1}$ ratio is examined and the results are shown in Fig. 3(e). The ratio of $ t_{0}/ t_{1}$ is restricted in the range of 0.0-3.5 because there is no new phase out of the range. The results reveal that the semimetal with hybrid nodes can only be found in the range of 0.0-1.0. At the phase transition point $ t_{0}/ t_{1}$ $=$ 1, the eigenvalues of the lowest three states are degenerate at the $\Gamma$ point as shown in Fig. S2(a). When $ t_{0}/ t_{1}$ $>$ 1, the degenerate bands at the $\Gamma$ point gap out and the charge-2 quadratic Weyl node around the $\Gamma$ point disappears. However, two Weyl points with charge -1 arise on the $\Gamma$-Z as shown in Fig. S2(b) corresponding to the situation of Fig. S1(c). There are three Weyl points in the systems, two of them with charge of -1 and one with charge of +2. The green and red lines in Fig. 3(e) show where the Weyl points with charge of -1 are, and they separate the trivial and nontrivial phases. It is significant to note that along with the evolution of the ratio, the bands at the A point remains fourfold degenerate with the Chern number of +2. In the range of the ratios, the first BZ is divided into three sections as shown in Fig. S3(b): the k$_z$ planes in the white zone have the Chern number of 0, which corresponds to normal insulator, whereas the other two zones are quantum anomalous Hall with Chern number of +1 or -1, respectively. However, only when the $ t_{0}/ t_{1}$ $<$ 1, the quantum anomalous Hall effect origins from the hybrid charge-2 semimetal phase.\\
\indent The above discussions mainly focus on the space group of $P4_{1}2_{1}2$. Similar conclusions can be produced in the space group $P4_{3}2_{1}2$. Moreover, the nontrivial hybrid single-pair charge-2 Weyl semimetal phase discussed above is not restricted to specific system. Generally, the new phase can be realized in spinless electronic materials, phononic or photonic crystals, acoustic metamaterials. In present work, we take phonons in a crystalline material as example to illustrate the realization of the nontrivial topological phase in realistic materials with the help of first-principles method.\\
\begin{figure}
	\includegraphics[trim={0.0in 0.0in 0.0in 0.0in},clip,width=5.4in]{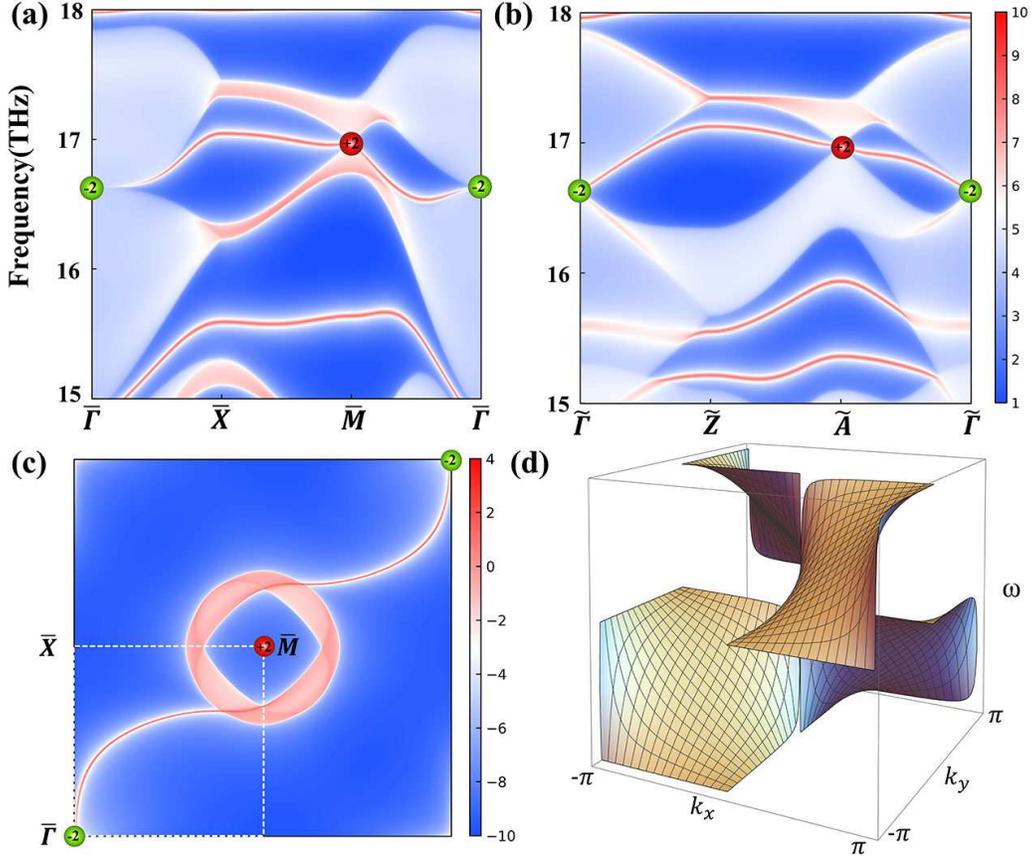}\\
	\caption{The surface LDOS of Na$_ 2$Zn$_ 2$O$_ 3$ in the (001) surface (a) and (010) surface (b). (c) The phonon surface arcs at 16.63 THz projected on the semi-infinite (001) surface. The green and red dots indicate the projections of Weyl points with $\mathcal{C} = +2$ and $\mathcal{C} = -2$, respectively. (d) The double-helicoid surface states are described by the Weierstrass elliptic function.}\label{fig5}
\end{figure}
\indent To concrete the hybrid single-pair charge-2 semimetal phase in term of realistic materials, through simply searching for the materials belong to above two nonsymmorphic space group ( $P4_12_12$ and $P4_{3}2_{1}2$), we find that the single-pair hybrid charge-2 semimetal states can be realized in the optic phonons of 3D Na$_ 2$Zn$_ 2$O$_ 3$. The material has been experimentally synthesized via solid state reaction method since 1996\cite{exp}. As shown in Fig. 4(a), it crystallizes in a tetragonal Bravais lattice with the space group $P4_32_12$. The optimized lattice constants are a = b = 6.174 $\AA$ and c = 9.397 $\AA$, which agree well with the experimental values\cite{exp}. The primitive cell has two inequivalent O atoms that occupy the Wyckoff positions 4a and 4b, respectively. Based on density functional perturbation theory\cite{dfpt}, we calculate the phonon spectra\cite{phonon} and density of states (DOS) of Na$_ 2$Zn$_ 2$O$_ 3$ with the Vienna ab initio simulation package (VASP)\cite{vasp}, the results are shown in Fig. 4(c). There are three different groups of isolated 4-band high-frequency (larger than 14 THz) optic modes and they are mainly contributed by the vibrations of O atoms. The zoom in phonon dispersions in the range of 14.9 to 18.0 THz, as shown in Fig. 4(d), indicate that a fourfold-degenerate point and a quadratic double-degenerate point present at the A and ${\Gamma}$ points, respectively, forming a single-pair of hybrid charge-2 nodes. The two bands forming the node around the $\Gamma$ point show quadratic dispersion in $k_z$=0 plane and  linear dispersion in other orientations, which is consistent with the above theoretical analysis. The Chern numbers of the Weyl points obtained with Wilson-loop method\cite{wilson,wanniertools} are denoted in Figs. 4(d), where green and red dots denote the Weyl points with Chern number of -2 and +2, respectively. Fig. 4(b) illustrates the distribution of the quadratic Weyl node $(\mathcal{C} = -2)$ at ${\Gamma }$ point (in the center of the BZ) and the charge-2 Dirac points $(\mathcal{C} = +2)$ at A point (at the corner of the BZ). The result is analogous to the theoretical conclusion above as shown in schematic Fig. 3(c).\\
\indent As mentioned above, one of the intriguing properties of the hybrid single-pair charge-2 semimetal phase is that the nontrivial surface states run through the entire projected FBZ of the system due to the unconventional Chern number distribution. To illustrate the topological characteristics throughout the whole BZ of Na$_ 2$Zn$_ 2$O$_ 3$, we compute the Chern number of horizontal two-dimensional planes on both sides of the Weyl point at the $\Gamma$ along $k_{z}$ orientation. The results show that the Chern numbers are 1 and -1 for $k_{z} > 0$ and $k_{z} < 0$, respectively, which shares the same nontrivial topological states with the case of Fig. 3(c). To further confirm the topological phonon properties of Na$_ 2$Zn$_ 2$O$_ 3$, we calculate the surface local density of states\cite{Green,iterative,wanniertools} (LDOS) on the (001) and (010) surfaces spanning the frequency range of 15 to 18 THz, the results are shown in Figs. 5(a) and 5(b). There are surface states connect the projected charge-2 nodes on the (001) and (010) surfaces further confirming the nontrivial properties of the single-pair charge-2 Weyl semimetal Na$_ 2$Zn$_ 2$O$_ 3$. The isofrequency surface for the frequency of 16.63 THz on the (001) surface exhibits chiral surface arcs rotating around the Weyl points, as shown in Fig. 5(c), that is the signature of nontrivial property for the material (the circles around the $\bar{M}$ derive from the energy difference between the two opposite charge Weyl points). The global picture of the surface states\cite{helicoid} also can be described by the Weierstrass elliptic function\cite{Chern_number2_2} and they are composed with two surface sheets wind around the two charge-2 Weyl points, as shown in Fig. 5(d).  The surface states are periodic in complex plane and have two singular points at (0, 0) and ($\pi$, $\pi$), which correspond to the positions of chiral charge +2 and -2, respectively.\\
\indent Using the theory of EBR, we propose that hybrid single-pair charge-2 Weyl semimetal can be forced in spinless systems with specific nonsymmorphic space groups of $P4_12_12$ and $P4_{3}2_{1}2$. Such new phase will provide an interesting platform to investigate the interaction between the two types of Weyl fermions with different dispersions. Moreover, the symmetry enforced Weyl node distribution produce nontrivial double-helicoid surface states that connect the projected nodes and run through the entire surface FBZ of the system, which is fascinating for future experimental detection and device applications. Although, in present work, we only take the simplified hybrid single-pair charge-2 optical phonon states in realistic material of Na$_2$Zn$_2$O$_3$ as an example, the novel phase will appear in the spinless electronic materials, phononic or photonic crystals, acoustic metamaterials as long as the nonsymmorphic symmetries are fulfilled. As for Na$_2$Zn$_2$O$_3$, its surface phonon states can be investigated experimentally using high resolution electron energy loss spectroscopy\cite{v1}, helium scattering\cite{v2}, or THz spectroscopy\cite{v3}. Moreover, the topological surface phonon states will improve electron-phonon interactions when the material form interface with other material that will benefit interfacial superconductivity.\\
\indent This work is supported by the National Natural Science Foundation of China (Grant No. 11804287, 11574260), Hunan Provincial Natural Science Foundation of China (2019JJ50577, 2021JJ30686),  and Hunan Provincial Innovation Foundation for Postgraduate (CX20190479).
\bibliography{references}
\bibliographystyle{apsrev}
\end{document}